\documentclass[aps,pra,twocolumn,amsmath,amssymb,nofootinbib,showpacs,superscriptaddress]{revtex4}

\usepackage[english]{babel}
\usepackage{latexsym}
\usepackage{graphics}
\usepackage{epsfig}
\usepackage{color}
\usepackage{bm}
\usepackage{amsmath}
\usepackage{amssymb}

\begin{document}

\title{{\color{black} Quantum Mechanical Stabilization of a Collapsing Bose-Bose Mixture}}

\author{D.~S.~Petrov}
\affiliation{Universit\'e Paris-Sud, CNRS, LPTMS, UMR8626, Orsay, F-91405, France}

\date{\today}

\begin{abstract}

According to the mean-field theory a condensed Bose-Bose mixture collapses when the interspecies attraction becomes stronger than the geometrical average of the intraspecies repulsions, $g_{12}^2>g_{11} g_{22}$. We show that instead of collapsing such a mixture gets into a dilute liquid-like droplet state stabilized by quantum fluctuations thus providing a direct manifestation of beyond mean-field effects. We study various properties of the droplet and find, in particular, that in a wide range of parameters its excitation spectrum lies entirely above the particle emission threshold. The droplet thus automatically evaporates itself to zero temperature, the property potentially interesting by itself and from the viewpoint of sympathetic cooling of other systems.

\end{abstract}


\maketitle

The mean-field and the first beyond mean-field contribution, the famous Lee-Huang-Yang (LHY) correction, to the ground state energy of a homogeneous weakly-repulsive Bose gas read \cite{LHY}
\begin{equation}\label{LHYscalar}
E/V=(gn^2/2)(1+128\sqrt{na^3}/15\sqrt{\pi}+...),
\end{equation}
where $n$ is the density and $a>0$ and $g=4\pi\hbar^2a/m$ are, respectively, the scattering length and coupling constant characterizing the interparticle interaction. The LHY correction originates from the zero-point motion of the Bogoliubov excitations and is thus intrinsically quantum. It is also universal in the sense that it depends only on the two-body scattering length and not on other parameters of the two-body or higher-order interactions. Quite naturally the experimental observation of this fundamental beyond mean-field effect came from the field of ultra-cold gases \cite{Altmeyer2007,Shin2008,Navon2010,Papp2008,Pollack2009,Navon2011}, where the gas parameter $na^3$ and, therefore, the relative contribution of the LHY term, can be enhanced by using Feshbach resonances \cite{ChinRMP}. Note however that the effect is perturbative; for $na^3\sim 1$ higher order terms and processes, in particular, three-body decay, come into play. A different situation is predicted for spinor gases where quantum fluctuations lift the degeneracy in the ground-state manifold \cite{Song2007,Turner2007} or lead to quantum mass acquisition \cite{Phuc2014}.

In this Letter we point out that in a Bose-Bose mixture the mean-field term and the LHY term depend on the inter- and intraspecies coupling constants in a different manner. Therefore, one can independently control them and make them comparable to each other without ever leaving the weakly-interacting regime. In particular, an interesting situation{\color{black}, impossible in the single-component case,} arises when the mean-field term, $\propto n^2$, is negative and the LHY one, $\propto n^{5/2}$, is positive. 
Because of its steeper density scaling the quantum LHY repulsion neutralizes the mean-field attraction and stabilizes the system against collapse. The mixture can then exist as a droplet in equilibrium with vacuum without any external trapping \cite{Skyrme}. This phenomenon {\color{black} naturally suggests a proof-of-principle experiment for observing the LHY quantum correction. The droplet can be prepared from currently available homo- and heteronuclear atomic mixtures by tuning the inter- and intraspecies scattering lengths into the unstable (from the mean-field viewpoint) region and by releasing the trap. We argue that several properties of the droplet are very unusual for ultracold gases and can have interesting implications. In particular, we predict that the droplet can have a peculiar excitation spectrum containing only a continuum part and very few or no discrete modes (usually associated with collective excitations: surface waves, breathing mode, etc.) This means that starting from an ordinary finite-temperature trapped condensed mixture one arrives at a macroscopic zero-temperature object; excitations corresponding to the continuum spectrum evaporate when the trap is switched off. The droplet can then be adiabatically manipulated (for instance, be trapped again) and used as a bath for sympathetic cooling of other systems. Another interesting property of the droplet is that its density can be strongly increased by going deeper into the unstable region and one can, in a controllable fashion, prepare an unprecedentally dense ultracold gas and study its optical properties, inelastic decay, etc.}

Let us choose the path-integral representation for the Bose-Bose mixture governed
by the action $S=\int d^3 {\bf r} dt\mathcal{L}[\Psi_1({\bf r},t),\Psi_1^*({\bf r},t),\Psi_2({\bf r},t),\Psi_2^*({\bf r},t)]$ with the Lagrangian density
\begin{equation}\label{LagrDensBare}
\mathcal{L}=\sum_i\Bigl[{\rm Re}(i\Psi_i^*\dot{\Psi}_i)-\frac{|\nabla_{\bf r}\Psi_i|^2}{2m_i}+\mu_i n_i\Bigr]-\sum_{ij} \frac{g_{ij}}{2}n_in_j,
\end{equation}
where $\mu_i$ is the chemical potential and $n_i({\bf r})=|\Psi_i({\bf r})|^2$ density of the $i$-th component, $g_{ii}=4\pi a_{ii}/m_{i}$ and $g_{12}=2\pi a_{12}/m_{\bf r}$ are, respectively, the intra- and interspecies coupling constants, $m_{\bf r}=m_1 m_2/(m_1+m_2)$, and we set $\hbar=1$. 

In the mean-field approximation the system is stable when the quadratic form $\sum_{ij} g_{ij}n_in_j$ is positive definite, which requires simultaneously $g_{11}>0$, $g_{22}>0$, and $g_{12}^2<g_{11}g_{22}$. The classical ground state is characterized by uniform condensate densities $n_i$ satisfying $\sum_j g_{ij}n_j=\mu_i$. In the weakly-interacting regime quantum fluctuations are weak and one can expand $\mathcal{L}$ up to quadratic terms in $\Psi'_i$ and $\Psi^{'*}_i$, where $\Psi'_i=\Psi_i-\sqrt{n_i}$, thus arriving at the Gaussian path integral. The validity condition for this approximation is simply $na^3\ll 1$ (here and after for qualitative estimates we omit subscripts assuming that the masses are of the same order of magnitude, $m_1\sim m_2 \sim m$, and the same holds for the densities, $n_1\sim n_2\sim n$, and scattering lengths, $a_{11}\sim a_{22}\sim |a_{12}|\sim a$). The equations of motion $\delta \mathcal{L}/\delta \Psi^{'*}_i=0$ and $\delta \mathcal{L}/\delta \Psi'_i=0$ give two Bogoliubov excitation branches \cite{Larsen,Oles}
\begin{equation}\label{Epm}
E_{k,\pm}=\sqrt{\frac{\omega_1^2+\omega_2^2}{2}\pm\sqrt{\frac{(\omega_1^2-\omega_2^2)^2}{4}+\frac{g_{12}^2n_1n_2k^4}{m_1m_2}}},
\end{equation}
where $\omega_i=\sqrt{g_{ii}n_ik^2/m_i+(k^2/2m_i)^2}$ are the Bogoliubov spectra for the individual components. 

The LHY correction is the zero-point energy corresponding to the Bogoliubov modes (\ref{Epm}).  The explicit expression for its density reads \cite{Larsen}
\begin{eqnarray}\label{LHYcontrib}
\frac{E_{\rm LHY}}{V}\!&\!=\!&\!\int\! \frac{d^3 k}{2(2\pi)^3} \left[ E_{+,k}+E_{-,k}-\frac{k^2}{2m_{\bf r}}-g_{11}n_1-g_{22}n_2\right.\nonumber\\
&+&\left.\frac{m_1g_{11}^2n_1^2+m_2g_{22}^2n_2^2+4m_{\bf r} g_{12}^2n_1n_2}{k^2}\right]
\end{eqnarray}   
and can be rewritten as
\begin{equation}\label{LHYrescaled}
E_{\rm LHY}/V=\frac{8}{15\pi^2}m_1^{3/2}(g_{11} n_1)^{5/2}f\left(\frac{m_2}{m_1},\frac{g_{12}^2}{g_{11}g_{22}},\frac{g_{22}n_2}{g_{11}n_1}\right),
\end{equation}
where $f>0$ is dimensionless. For $m_2=m_1$ and for trivial cases, such as $n_i=0$ or $g_{12}=0$, the two excitation branches~(\ref{Epm}) have the form of the usual single-component Bogoliubov spectra and the integral in Eq.~(\ref{LHYcontrib}) is analytic. In particular, for equal masses $f(1,x,y)= \sum_{\pm}(1+y\pm\sqrt{(1-y)^2+4xy})^{5/2}/4\sqrt{2}$. In any case, the main contribution to the integral in Eq.~(\ref{LHYcontrib}) comes from momenta of order $1/\xi_{\rm h}=\sqrt{mgn}$.

Let us now introduce $\delta g=g_{12}+\sqrt{g_{11}g_{22}}$ and discuss the unstable regime when $\delta g$ is negative but small compared to $g_{11}>0$ and $g_{22}>0$. The mechanical instability can be understood by diagonalizing the mean-field term  $\sum_{i,j=1,2}g_{ij}n_in_j/2=\sum_{\pm}\lambda_{\pm}n_{\pm}^2$, where $\lambda_{-}\approx \delta g\sqrt{g_{11}g_{22}}/(g_{11}+g_{22})$ is negative and small, $\lambda_{+}\approx (g_{11}+g_{22})/2$, $n_{-}=(n_1\sqrt{g_{22}}+n_2\sqrt{g_{11}})/\sqrt{g_{11}+g_{22}}$, and $n_{+}=(n_1\sqrt{g_{11}}-n_2\sqrt{g_{22}})/\sqrt{g_{11}+g_{22}}$. It is energetically favorable to maximize $n_{-}^2$ and minimize $n_{+}^2$, i.e., increase both densities while preserving the ratio $n_2/n_1={\rm const}=\sqrt{g_{11}/g_{22}}$. {\color{black} The instability also manifests itself in the fact that $E_{k,-}$ becomes complex for small momenta $k\sim \sqrt{m|\delta g|n}$. However, for $k\sim 1/\xi_{\rm h} \gg \sqrt{m|\delta g|n}$, i.e., in the region mostly contributing to the LHY term, both modes $E_{k,-}$ and $E_{k,+}$ are insensitive to small variations of $\delta g$ and, in particular, to its sign. Note that the global increase in densities leads to the hardening of these modes and to the growth of the corresponding zero-point energy $\propto n^{5/2}$ which is faster than the mean-field energy gain $\propto n^2$. One can thus say that the long-wave-length instability is cured by quantum-mechanical fluctuations at shorter wave lengths.}

More formally, let us introduce a characteristic momentum $p_c$ such that $\sqrt{m|\delta g|n} \ll p_c \ll 1/\xi_{\rm h}$. One can then obtain an effective low-energy theory by integrating out the modes with $p>p_c$ from the initial problem. The result is that the effective Lagrangian density (for the low-momentum part of $\Psi_i$) is given by Eq.~(\ref{LagrDensBare}) minus the zero-point energy of the high-energy modes. The latter is given by Eq.~(\ref{LHYcontrib}) or (\ref{LHYrescaled}), where one can set $g_{12}^2=g_{11}g_{22}$ neglecting small finite-$\delta g$ corrections and extending the integration interval $p_c<k<\infty$ to $0<k<\infty$ by using the fact that the low-momentum contribution to Eq.~(\ref{LHYcontrib}) is negligible.

As a result of the competition between the attractive mean-field term $\propto n^2$ and repulsive LHY term $\propto n^{5/2}$ the mixture can exist at finite density without any trapping, i.e., in equilibrium with vacuum. Note that the LHY term competes with the {\color{black} attractive term} $\lambda_{-}n_{-}^2$ but it is still much too weak compared to $\lambda_{+}n_{+}^2$. This locks the ratio of the equilibrium densities $n_2^{(0)}/n_1^{(0)}=\sqrt{g_{11}/g_{22}}$. Then, omitting the numerical prefactors, the structure of the interaction part of the energy functional reads $gn_{+}^2-|\delta g|n_{-}^2+m^{3/2}(gn_{-})^{5/2}$ leading to the equilibrium value $n_{-}^{(0)}\sim \delta g^2/m^3g^5\sim (\delta g/g)^2/a^3$. This means that the system remains weakly interacting (gaseous parameter $na^3\ll 1$) under the condition $(\delta g/g)^2\ll 1$. More quantitatively, we obtain
\begin{equation}\label{Density}
n_1^{(0)}=\frac{25\pi}{1024}\frac{1}{f^2(m_2/m_1,1,\sqrt{g_{22}/g_{11}})}\frac{1}{a_{11}^3}\frac{\delta g^2}{g_{11}g_{22}},
%/f^{2}\left(\frac{m_2}{m_1},1,\sqrt{\frac{g_{22}}{g_{11}}}\right)\frac{\delta g^2}{g_{11}g_{22}}\frac{1}{a_{11}^3}
%\frac{1}{f^2(m_2/m_1,1,\sqrt{g_{22}/g_{11}})}.
%\left(1+\sqrt{\frac{g_{22}}{g_{11}}}\right)^{-5}.
\end{equation}
and explicitly for equal masses
\begin{equation}\label{DensEqMass}
n_i^{(0)}|_{m_1=m_2}=\frac{25\pi}{1024}\frac{(a_{12}+\sqrt{a_{11}a_{22}})^2}{a_{11}a_{22}\sqrt{a_{ii}}(\sqrt{a_{11}}+\sqrt{a_{22}})^5}.
\end{equation}

For finite particle numbers the system is in the droplet state. In order to study its density profile and low-lying excitations we simplify the problem by setting $\Psi_i({\bf r},t)=\sqrt{n_i^{(0)}}\phi({\bf r},t)$, where $\phi({\bf r},t)$ is a scalar wave function \cite{remPhase}. This assumption neglects possible relative motion of the components, i.e., {\color{black} energy-expensive fluctuations of $n_{+}$}, and more exotic situations such as, for example, vortex excitations characterized by different charges in the two components. However, it is justified for the ground state of the droplet and the {\color{black} low-energy} part of its spectrum. With this reservation, the effective theory for the field $\phi({\bf r},t)$ is governed by the action {\color{black}
\begin{equation}\label{Lagrphi}
S=\eta\int d^3 \tilde{\bf r} d\tilde{t}[{\rm Re}(i\phi^*\partial_{\tilde{t}} \phi)-\tilde\epsilon(\phi,\phi^*)+\tilde{\mu}|\phi|^2],
\end{equation}
where $\eta=(2/3)|\delta g|n_1^{(0)}n_2^{(0)}\xi^3\tau$ and $\tilde\epsilon(\phi,\phi^*)=|\nabla_{\tilde{\bf r}} \phi|^2/2-3|\phi|^4/2+|\phi|^5$ is the rescaled energy density.} We have introduced the rescaled coordinate $\tilde{\bf r}={\bf r}/\xi$ and time $\tilde{t}=t/\tau$, where 
\begin{equation}\label{xitau}
\xi=\sqrt{\frac{3}{2}\frac{\sqrt{g_{22}}/m_1+\sqrt{g_{11}}/m_2}{|\delta g|\sqrt{g_{11}}n_1^{(0)}}},\;\tau=\frac{3}{2}\frac{\sqrt{g_{11}}+\sqrt{g_{22}}}{|\delta g|\sqrt{g_{11}}n_1^{(0)}}.
\end{equation}
In what follows we denote rescaled quantities by tilde.

Since the coefficient $\eta$ is large, $\eta\sim (g/|\delta g|)^{5/2}\gg 1$, the problem (\ref{Lagrphi}) can be treated (quasi)classically. The equation of motion reads
\begin{equation}\label{EqMot}
i\partial_{\tilde{t}}\phi = (-\nabla_{\tilde{\bf r}}^2/2-3|\phi|^2+5|\phi|^3/2-\tilde{\mu})\phi
\end{equation}
and the Gross-Pitaevskii (GP) equation for the ground state is obtained by setting $\phi(\tilde{\bf r},\tilde{t})=\phi_0(\tilde{\bf r})$ \cite{remMGP}. The chemical potential $\tilde{\mu}$ is fixed by the normalization condition $\tilde{N}=\int d^3\tilde{r}|\phi_0|^2$, where $\tilde{N}$ is related to the number of particles of the $i$-th component by $N_i=n_i^{(0)}\xi^3\tilde{N}$. The uniform solution for infinite $\tilde{N}$ corresponds to $\phi_0\equiv 1$ and $\tilde{\mu}=-1/2$.

For large but finite $\tilde{N}$ the ground state is a spherical droplet of large radius $\tilde{R}\approx (3\tilde{N}/4\pi)^{1/3}$ with approximately unit bulk (saturation) density. Near the surface, if we denote by $\tilde{x}$ the coordinate normal to it, the wave function $\phi_0(\tilde{x})$ {\color{black} satisfies the one-dimensional GP equation of the form $d^2\phi_0(\tilde{x})/d\tilde{x}^2=-d U(\phi_0)/d\phi_0$, which describes the classical motion of a particle with coordinate $\phi_0$ and time $\tilde{x}$ in the potential $U(\phi_0)$ \cite{1Dsoliton}. The integral of motion $(d\phi_0/d\tilde{x})^2/2+U(\phi_0)=0$ can be integrated again resulting in the implicit formula for $\phi_0(\tilde{x})$,}
\begin{equation}\label{Interface}
\tilde{x}(\phi_0)=\frac{1}{\sqrt{3}}\ln\frac{\sqrt{3}-\sqrt{1+2\phi_0}}{\sqrt{3}+\sqrt{1+2\phi_0}}+\ln\frac{\sqrt{1+2\phi_0}+1}{\sqrt{1+2\phi_0}-1},
\end{equation}
from which we see that the surface thickness is of order $\xi$ (in initial units). That the corresponding momentum scale $1/\xi$ is much smaller than $p_c$ a posteriori justifies the applicability of the low-energy effective theory (\ref{Lagrphi}). {\color{black} The surface tension equals $\tilde{\sigma}=\int d\tilde{x}[\tilde\epsilon(\phi_0,\phi^*_0)-\tilde\mu|\phi_0|^2]=3(1+\sqrt{3})/35$ and we} obtain the spectrum of the droplet's surface modes (ripplons) {\color{black} valid in the limit of large $\tilde{N}$} \cite{Barranco}
\begin{equation}\label{Ripplons}
\tilde{\omega}_l=\sqrt{(4\pi/3)l(l-1)(l+2)\tilde{\sigma}/\tilde{N}},
\end{equation}
where {\color{black} $l>0$} is the angular momentum. {\color{black} The dipolar mode ($l=1$) can be regarded as a center-of-mass displacement of the droplet, therefore $\tilde{\omega}_1\equiv 0$}.

\begin{figure}
\centerline{\includegraphics[width=0.95\hsize,clip,angle=0]{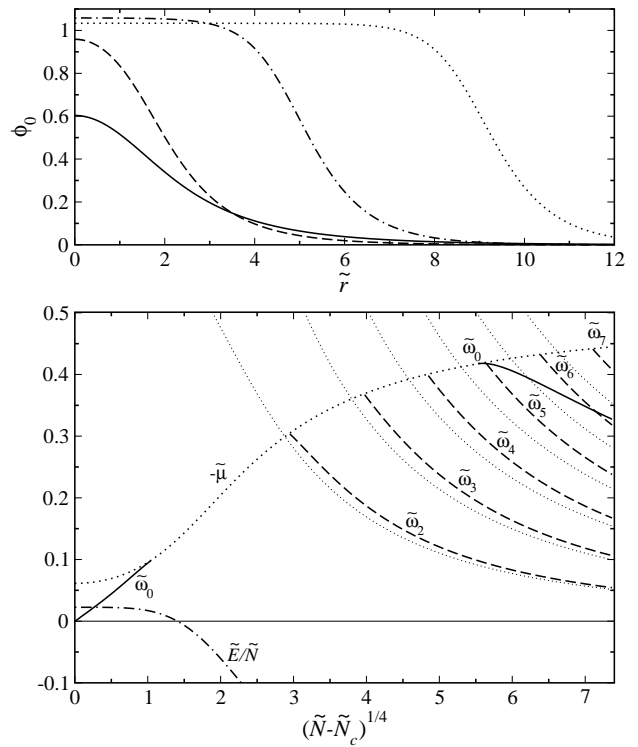}}
\caption{Upper panel: the droplet wave function vs radial coordinate for $\tilde{N}=\tilde{N}_c\approx 18.65$ (solid), $\tilde{N}=30$ (dashed), $\tilde{N}=500$ (dash dotted), and $\tilde{N}=3000$ (dotted). Lower panel: the rescaled energy per particle $\tilde{E}/\tilde{N}$ (dash-dotted), particle emission threshold $-\tilde{\mu}$ (thick dotted), monopole mode frequency $\tilde{\omega}_0$ (solid), frequencies of higher angular momentum modes $\tilde{\omega}_l$ (dashed), and the corresponding surface-mode approximation Eq.~(\ref{Ripplons}) (thin dotted) versus $(\tilde{N}-\tilde{N}_c)^{1/4}$.}
\label{fig:Specra}
\end{figure}

In general, when $\tilde{N}$ is not large, the size of the droplet is comparable to the surface thickness. In this case, we calculate the wave function $\phi_0$ (see the upper panel in Fig.~1) and determine the chemical potential $\tilde{\mu}$ {\color{black} of the droplet and its energy $\tilde{E}=\int d^3\tilde{\bf r}\tilde\epsilon(\phi_0,\phi^*_0)$} numerically. We also study its small-amplitude excitations by solving the Bogoliubov-de Gennes equations obtained by linearizing Eq.~(\ref{EqMot}) with respect to small $\phi(\tilde{\bf r},\tilde{t})-\phi_0(\tilde{\bf r})$. We find that with decreasing $\tilde{N}$ the droplet becomes metastable ($\tilde{E}>0$) for $\tilde{N}<22.55$ and unstable for $\tilde{N}<\tilde{N}_c\approx 18.65$. This effect can be understood by the following qualitative arguments. For a droplet of size $\tilde{R}$ the energy is composed of the kinetic term $\propto \tilde{N}/\tilde{R}^2$, two-body interaction $\propto -\tilde{N}^2/\tilde{R}^3$, and LHY contribution $\propto \tilde{N}^{5/2}/\tilde{R}^{9/2}$. {\color{black} With decreasing $\tilde{N}$} the minimum of $\tilde{E}(\tilde{R})$ first becomes metastable (the global ground state corresponds to $\tilde{R}\rightarrow \infty$) and then disappears completely. We should mention the analogy of this situation with the problem of a harmonically trapped scalar Bose-Einstein condensate with attractive two-body interactions (see \cite{RMPBEC} for review). In that case the LHY correction is negligible and the stabilizing role is played by the potential energy $\propto \tilde{N}\tilde{R}^2$. There is also a metastable minimum which becomes unstable (in this case with increasing $\tilde{N}$) as a result of the interplay of the kinetic, interaction, and potential energies.

In the lower panel of Fig.~1 we show the quantity $\tilde{E}/\tilde{N}$ (dash-dotted line), particle emission threshold $-\tilde{\mu}$ (thick dotted line) {\color{black} which separates the discrete and continuum parts of the spectrum}, frequency of the monopole mode $\tilde{\omega}_0$ (solid line), and frequencies of higher angular momentum modes $\tilde{\omega}_l$ (dashed lines) as functions of $(\tilde{N}-\tilde{N}_c)^{1/4}$. The thin dotted lines extending above the particle emission threshold represent the result of Eq.~(\ref{Ripplons}). All excitation modes cross the threshold for sufficiently small $\tilde{N}$. Only the monopole mode %crosses the threshold at $\tilde{N}\approx 954$ [$(\tilde{N}-\tilde{N}_c)^{1/4}\approx 5.5$], then 
reenters at $\tilde{N}\approx 20.1$. % and its frequency near the instability behaves as $\tilde{\omega}_0\propto (\tilde{N}-\tilde{N}_c)^{1/4}$ {\color{black} as we will discuss below}. 
{\color{black} Remarkably, in the interval $20.1<\tilde{N}<94.2$ there are no modes below $-\mu$ and, therefore, exciting the droplet is equivalent to spilling of particles or, more generally, to breaking the droplet into smaller pieces. We thus deal with an automatically evaporating object which is also macroscopic since the actual particle numbers are large, $N_i=n_i^{(0)}\xi^3 \tilde{N}\sim (g/|\delta g|)^{5/2}\tilde{N}\gg \tilde{N}$.}

%Speaking of the hard modes they are all ``unbound''; the lowest frequency can be estimated as $E_{1/\xi,+}$, which, in the rescaled units, gives $\tilde{\omega}_{\rm hard}\sim \xi/\xi_{\rm hard}\sim \sqrt{g/|\delta g|}\gg 1$. For completeness, let us also mention here that the ratio $N_2/N_1$ is not locked down to the $1/N$ accuracy. However, there is a critical imbalance above which the excess particles are no longer bound to the droplet. Adding, say, $\delta N_1$ excess particles to the first component makes $\tilde{N}$ proportionally larger leading to a linear gain in energy $\delta \tilde{E}=\tilde{\mu} \delta \tilde{N}\propto -\delta N_1$. On the other hand there is a quadratic penalty coming from the term $\lambda_{+}n_{+}^2\propto \delta N_1^2$. One can show that the excess particles are spilled out when $\delta N_1/N_1$ becomes larger than  a critical value $\sim |\delta g|/g$. This imbalance introduces relative corrections of order $|\delta g|/g\ll 1$ to the results obtained for $n_{+}\equiv 0$.

{\color{black} Let us now discuss excitations involving relative motion of the components with finite $n_{+}$, neglected in deriving Eq.~(\ref{Lagrphi}). In the homogeneous case these modes correspond to $E_{+,k}$ and, for the droplet of size $\sim 1/\xi$, we can estimate the lowest frequency to be $E_{+,1/\xi}$. In rescaled units this quantity is approximately $\xi/\xi_{\rm h}\sim \sqrt{g/|\delta g|}\gg 1$, i.e., all such modes are in the continuum. It is also possible to have a nonzero $n_{+}$ without exciting the relative motion. Indeed, starting from the configuration $N_2/N_1=\sqrt{g_{11}/g_{22}}$ and adding, say, $\delta N_1$ excess particles to the first component makes $\tilde{N}$ proportionally larger leading to a linear gain in energy $(\eta/\tau)\tilde{\mu} \delta \tilde{N}\sim -|\delta g| n \delta N_1$. However, due to the quadratic penalty coming from the term $\lambda_{+}n_{+}^2 R^3\sim g\delta N_1^2/R^3$ (here $R=\xi\tilde{R}$ is the droplet radius) one can increase $\delta N_1/N_1$ only up to a critical value $\sim |\delta g|/g\ll 1$. Beyond this point the excess particles no longer bind to the droplet.}

%For completeness, let us also mention here that the ratio $N_2/N_1$ is not locked down to the $1/N$ accuracy. However, there is a critical imbalance above which the excess particles are no longer bound to the droplet. Adding, say, $\delta N_1$ excess particles to the first component makes $\tilde{N}$ proportionally larger leading to a linear gain in energy $\delta \tilde{E}=\tilde{\mu} \delta \tilde{N}\propto -\delta N_1$. On the other hand there is a quadratic penalty coming from the term $\lambda_{+}n_{+}^2\propto \delta N_1^2$. One can show that the excess particles are spilled out when $\delta N_1/N_1$ becomes larger than  a critical value $\sim |\delta g|/g$. This imbalance introduces relative corrections of order $|\delta g|/g\ll 1$ to the results obtained for $n_{+}\equiv 0$.

%In the vicinity of the instability point the droplet can decay by quantum fluctuations. {\color{black} The mechanism is similar to the quantum tunneling of an attractive spinless trapped Bose gas towards collapse \cite{Ueda}, although in our case the droplet actually becomes unbound and expands to infinity. Both cases are associated with a softening of the breathing (monopole) mode and its tunneling under a barrier.}

In the vicinity of the instability point the droplet can decay by quantum fluctuations {\color{black} similarly to the quantum tunneling of an attractive scalar trapped Bose gas towards collapse \cite{Shuryak,Ueda}. Although in our case we are dealing with an expansion to infinity rather than collapse, in both cases the mechanism is associated with a softening of the breathing (monopole) mode and its tunneling under a barrier.} In order to estimate the corresponding lifetime we first note that near the instability point the chemical potential behaves as $\tilde{\mu}\approx \tilde{\mu}_c-\sqrt{2(\tilde{N}-\tilde{N}_c)/\tilde{N}''}$ {\color{black} \cite{remOtherBranch}}, where we find numerically $\tilde{\mu}_c\approx 0.061$ and $\tilde{N}''=\partial^2\tilde{N}(\tilde{\mu}_c)/\partial \tilde{\mu}^2\approx 2190$. Let us parameterize $\phi_0$ by the chemical potential and introduce the derivative $\delta \tilde{n}(\tilde{r})=|\phi_0(\tilde{r})|\partial |\phi_0(\tilde{r})|/\partial \tilde{\mu}$ taken exactly at the critical point. An excitation ``along'' $\delta \tilde{n}(\tilde{r})$ does not change $\tilde{N}$ and corresponds to the monopole mode, the frequency of which vanishes at the instability point. Close to this point the tunneling path goes along this soft degree of freedom and we introduce the corresponding coordinate $\lambda$ by writing the density as $\tilde{n}(\tilde{r},\tilde{t})=|\phi_0(\tilde{r})|^2+\lambda(\tilde{t})\delta \tilde{n}(\tilde{r})$ and the wave function as $\phi(\tilde{r},\tilde{t})=\sqrt{\tilde{n}(\tilde{r},\tilde{t})}\exp[i\theta(\tilde{r},\tilde{t})]$. Here the phase $\theta$ is determined by minimizing (\ref{Lagrphi}) for a given trajectory $\lambda(\tilde{t})$, which is equivalent to solving the continuity equation $-\dot{\tilde{n}}=\nabla_{\tilde{\bf r}}(\tilde{n}\nabla_{\tilde{\bf r}}\theta)$ with respect to $\theta$. In this manner we obtain the effectively one-dimensional problem with the action 
\begin{equation}\label{Instanton}
S=\eta\int d\tilde{t} [m_{\rm eff}\dot{\lambda}^2/2-U_{\rm eff}(\lambda)],
\end{equation} 
where the effective mass $m_{\rm eff}=\pi \int_0^\infty [\int_0^{\tilde{r}} \tilde{r}^{'2}\delta\tilde{n}(\tilde{r}')d\tilde{r}']^2/[\tilde{r}^2\tilde{n}_c(\tilde{r})]d\tilde{r}\approx 2240$ [here $\tilde{n}_c(\tilde{r})$ is the density profile at the instability point] and effective potential $U_{\rm eff}(\lambda)=(\tilde{N}''/8)[(\tilde{\mu}_c-\tilde{\mu})\lambda^2-\lambda^3/6]$. The tunneling rate out of such a quadratic-plus-cubic potential in the quasiclassical regime equals $\Gamma=A\exp(-S^{\rm B})$ \cite{Ueda}, where the bounce exponent $S^{\rm B}=(3/5)2^{21/4}\eta m_{\rm eff}^{1/2}\tilde{N}^{''-3/4}(\tilde{N}-\tilde{N}_c)^{5/4}\approx 3.38\eta (\tilde{N}-\tilde{N}_c)^{5/4}$, $A\tau=\tilde{\omega}_0\sqrt{15S^{\rm B}/2\pi}$, and the monopole frequency $\tilde{\omega}_0=[\tilde{N}''(\tilde{N}-\tilde{N}_c)/8m_{\rm eff}^2]^{1/4}$ {\color{black} (see Fig.~\ref{fig:Specra})}. Thus, for the observation of the macroscopic tunneling, due to the exponential dependence of $\Gamma$, the most relevant region of parameters is $\tilde{N}-\tilde{N}_c\sim \eta^{-4/5}\sim (\delta g/g)^2$.

Let us now discuss possibilities of having such droplets in experiments with ultracold gases. A very promising candidate is the mixture of the second and third lowest hyperfine states, $F=1,m_F=0$ (state 1) and $F=1,m_F=-1$ (state 2), of $^{39}$K. The corresponding scattering lengths $a_{ij}$ as functions of the magnetic field have been studied theoretically \cite{DErricoNJP2007,Lysebo} and experimentally \cite{DErricoNJP2007}. In particular, the condition $a_{12}=-\sqrt{a_{11}a_{22}}$ is satisfied at $B_0\approx 56.77$G where $a_{11}\approx 84.3 a_{\rm Bohr}$, $a_{22}\approx 33.5 a_{\rm Bohr}$, and the quantity $\partial(a_{12}+\sqrt{a_{11}a_{22}})/\partial B\approx 12.09 a_{\rm Bohr}/{\rm G}$. For example, for $B-B_0=-250$mG the saturation densities equal  $n_1^{(0)}\approx 3.3\times 10^{14}$cm$^{-3}$ and $n_2^{(0)}\approx 4.9\times 10^{14}$cm$^{-3}$, length scale $\xi\approx 1.96\mu$m, and time scale $\tau \approx 2.4$ms. Then, $\tilde{N}=30$ corresponds to $N_1=0.75\times 10^{5}$, $N_2=1.1 \times 10^{5}$, and the droplet wave function is given by the dashed line in Fig.~1 (upper panel). We can compare the characteristic time scale $\tau$ with the lifetime $\tau_{\rm life}\sim K_3^{-1}(n_1+n_2)^{-2}\approx 150$ms, where we use the three-body recombination rate constant $K_3=10^{-29}$cm$^6$/s \cite{ZaccantiNatPhys2009}. It is also useful to note that $\tau$ scales as $|\delta g|^{-3}$ whereas $\tau_{\rm life}\propto |\delta g|^{-4}$, i.e., there is still room for increasing $|\delta g|$ while keeping $\tau<\tau_{\rm life}$. Thus, one can have very dilute and slow droplets as well as dense and fast ones. Another interesting candidate is the dual-species mixture of $^{87}$Rb and $^{39}$K both in the $F=1,m_F=-1$ hyperfine state. This mixture is characterized by positive intraspecies interactions close to an interspecies Feshbach resonance at $B=117.56$G, which has recently been explored experimentally \cite{Wacker}. In any case, unless we are dealing with the microgravity environment \cite{Microgravity}, the droplet has to be levitated against gravity. This can be achieved by using, for example, an optical potential with a vertical gradient. Our theory can be modified to include a more general external confinement, say, harmonic. However, the self-binding and related properties make sense only if the system is unconfined at least in one (horizontal) direction.

We thank P. Schuck and S. Stringari for fruitful discussions and acknowledge support by the IFRAF Institute. The research leading to these results received funding from the European Research Council (FR7/2007-2013 Grant Agreement No. 341197).


\begin{thebibliography}{99}
\bibitem{remPhase} A possible (constant) phase difference between $\Psi_1$ and $\Psi_2$ plays no role as can be seen from Eq.~(\ref{LagrDensBare}).
\bibitem{remOtherBranch} {\color{black} In fact, the GP equation has a solution also for the other branch of the square root, $\tilde{\mu}_c+\sqrt{2(\tilde{N}-\tilde{N}_c)/\tilde{N}''}$, corresponding to a local maximum of the energy functional.}
\end{thebibliography}
\end{document}